\documentclass[12pt]{article}
\usepackage{graphics}
\usepackage{epsfig} 
\usepackage{hyperref}    
\usepackage{color} 
\usepackage{graphics}
\usepackage{graphicx}
\usepackage{latexsym,amssymb}
\title{\bf Lateral phase separation of confined membranes }
\author{Mesfin Asfaw \\
     Molecule and Life Nonlinear Sciences Laboratory, Research \\Institute 
     for Electronic Science, Hokkaido University, Kita 20 Nishi 10, Kita-ku, \\Sapporo 001-0020, Japan}

\begin{document} 
\maketitle
\date{Received: date / Revised version: date}

\abstract{ We consider membranes interacting via short, intermediate and long stickers. The effects of the intermediate stickers on the lateral phase separation of the membranes are studied via mean-field approximation. The critical potential depth of the stickers increases in the presence of the intermediate sticker. The lateral phase separation of the membrane thus  suppressed by the intermediate stickers. Considering membranes interacting with short and long stickers,  the effect of confinement on the phase behavior of the membranes is also investigated analytically.}
%
\maketitle

\section{\bf Introduction}

Membranes serve  a number of general functions in our cells and tissues. They separate cells and cell compartments \cite{h1,h2}.
They act as barriers as their lipid core is permeable to water and to small hydrocarbon molecules and relatively impermeable
to macromolecules and polar molecules.  In addition, membrane proteins facilitate or assist the transport of ions and macromolecules  into and 
out of the cells. Since biomembranes play a vital role in biological process, they have attracted  significant experimental  
 and theoretical  interest  \cite{h19,h20,h21,t1}.  The physics of membrane adhesion is one of the aspects that  has got 
  considerable attention \cite{h3,h4,h5,h6,h7,h8,h9,h11,hh12,hh13,hh14}.  Recent  experimental  realization  \cite{h12}  uncovers  the formation of 
 domains between  short and long receptors of T cells. The dynamics of adhesion induced phase separation 
  was also  considered theoretically  in the works \cite{h13,h14} and its detailed equilibrium studies were reported in the previous studies  \cite{h16, h17,h18}. 
   All these theoretical and experimental investigations have significantly  contributed to  the basic understanding  of  the physics of membranes. 
   More recently,  we presented  a theoretical study for adhesion-induced lateral phase separation of  membranes with short stickers, 
   long stickers and repellers confined between two hard walls \cite{h22}. The effects of confinement and repellers on lateral 
   phase separation were  investigated. It is found  that the sticker critical potential depth tends to 
   increase as the distance between the hard walls decreases which  suggests confinement-induced or force-induced mixing of stickers.

Despite the success to understand the thermodynamic properties of these non-homogeneous membranes, more efforts are needed to get a complete picture of the physics of membranes due to the fact that membranes contain large number of
macromolecules which are organized in complex fashion. Most of the recent works considered   membranes interacting with one or two types of stickers (receptor/ligand pairs). However the adhesive molecules may have different resting length;  the experimental investigations of T cells unveiled two or more  receptor-ligand pairs being involved in the binding process \cite{hh12}.
Another crucial issue on the adhesion induced lateral phase separation is the effect of confinement. Cell adhesion often occurs in the presence of external force field due to external flow \cite{h22}.
The aim of this work is to investigate the phase behavior of confined membranes interacting with more than one types of stickers. First  
we investigate the phase behavior of membranes with short and long stickers analytically which 
can be taken as an independent check of the results found
in the previous works \cite{h18,h22}. We then extend  our previous studies  by considering        membranes interacting with  three  adhesive molecules.

The rest of the work is structured as follows.  In section II, we introduce our model.  In section III,  we present the mean field approximation
 while section IV  and V deal with membranes interacting via two  and three types of stickers, respectively. Section VI deals with summary and conclusion.

\section{The model }

We consider non-homogenous membrane that interacts via receptor/ligands (stickers) of different characteristic length as shown in Fig. 1. 
For simplicity, a discretized membrane (lattice gas model \cite{h8, h18})    with lattice constant $a$ is considered.  The separation field $l$
 separates the membrane from the  membrane (substrate) while the occupation number $n_{i}=0$ shows no sticker present at the lattice site $i$ 
 while  $n_{i}=1$, $2$, $3$... indicates   the presence of stickers 1, stickers 2, stickers 3... at lattice site $i$,  respectively (see Fig. 1).
\begin{figure}[h] 
\begin{center}
\epsfig{file=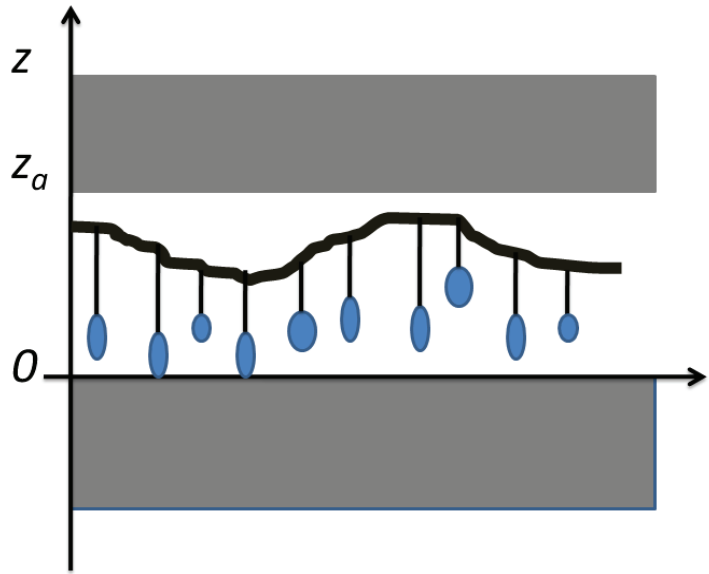,width=10cm} 
\caption{Schematics diagram 
of membrane with stickers of different size near to a substrate. $z$  denotes the local separation field that separates the membrane from the  substrate.}
\end{center}
\end{figure}

     Introducing the rescaled fields $z=(l/a)\sqrt{k/T}$ and ${ V}=V/T$ reduces the number of independent parameters. 
For tensionless non-homogenous multi-component membranes, 
the Hamiltonian of the model is given by 
         \begin{eqnarray}
         H[z,n]&=&H_{el}[z]+   \sum_{i} \delta_{1,n_i}(V_{1}(z_i)-\mu_{1})+ \nonumber \\&&
       \sum_{i} \delta_{2,n_i}(V_{2}(z_i)-\mu_{2})+\sum_{i}\delta_{3,n_i}(V_{3}(z_i)-\mu_{3})\nonumber \\&&
           +\sum_{i}\delta_{4,n_i}(V_{4}(z_i)-\mu_{4})+...   
         \end{eqnarray}
with the elastic term
 \begin{equation}
         H_{el}[z]= \sum_{i}{ \kappa \over 2a^2}(\Delta_{d}z_i)^2
         \end{equation}
	where        
$H_{el}[{z}]$   denotes the discretized bending energy of the membranes with an effective bending rigidity $\kappa $.   Typically, $\kappa = 10-20k_{B}T$. 
The discretized Laplacian $\Delta_d $ is given by $
        \Delta_{d}z_i=z_{i1}+z_{i2}+z_{i3}+z_{i4}-4z_{i}$. 
         $V_{1}({\it z}_{i})$,  $V_{2}({\it z}_{i})$ and  $V_{3}({\it z}_{i})$     denote the  potentials for stickers 1, stickers 2, and stickers 3, respectively while  $\mu_{1}$, $\mu_{2}$  and
         $\mu_{3}$ denote the chemical potentials for the stickers 1, stickers 2 and stickers 3, respectively. 
      The partition function $Z$ can be written as
       \begin{equation}
         Z=\left[\prod_{i} \int_{-\infty}^{\infty }dz_{i}\right]\left[\prod_{i} \sum_{n_{i}=0,1,2} \exp \left[{-H[z,n]} \right]
         \right].
           \end{equation}
           Tracing out the sticker degrees of freedom one gets
       \begin{eqnarray}   
           Z&=&\prod_{i} \int_{-\infty}^{\infty }dz_{i}\left[\exp[{-H_{el}(z)}\right] \nonumber \\ &&
(1+\exp \left[{-V_{1}(z_i)+\mu_{1}}\right]
           +  \exp \left[{-V_{2}(z_i)+\mu_{2}}\right]+  \nonumber \\ && \exp \left[{-V_{3}(z_i)+\mu_{3}}\right]+...).
           \end{eqnarray}
           Rearranging  terms,  Equation (4) can be written as
            \begin{eqnarray}
            Z&=&(1+\exp[{\mu_{1}}]+\exp[{\mu_{2}}]+\exp[{\mu_{2}}]+\exp[{\mu_{3}}]...)^N  \nonumber \\&& \prod_{i}\int_{-\infty}^{\infty }dz_{i}
            \exp \left[{-H_{el}(z)+ \sum_{i} [V^{eff}(z_i)]}\right].
         \end{eqnarray}
         The effective potential $V^{eff}(z)$ is given by  
         \begin{eqnarray}
         V^{eff}(z)&=&-\ln[(1+\exp[{-V_{1}(z)+\mu_{1}}]+
           \exp[{-V_{2}(z)+\mu_{2}}]+  \nonumber  \\ &&
	   \exp[{-V_{3}(z)+\mu_{3}}]+...)/ \xi_{0} ]
           \end{eqnarray}
	    where $\xi_{0}=1+\exp[{\mu_{1}}]+\exp[{\mu_{2}}]+\exp[{\mu_{3}}]+...$. Integrating out the stickers  degrees of freedom  leads  to a homogeneous membrane with an effective potential  which is  given by Eq. (6). The sticker's potential is modeled as square-well potential \cite{h8, h18})    
         $ V_1=
              U_{1}$ for $z_{1}< z_{i} <  z_{2}$,  $V_2=
              U_{2}$ for  $z_{3}< z_{i}  < z_{4}$, $V_3=
              U_{3}$ for  $z_{5}< z_{i}  < z_{6}$ and $V_{1}$, $V_{2}$, $V_{3}$...=0 otherwise. The 
                effective potential  (6) then can be  rewritten as   
${ V}^{eff} =\infty$  for $z<0$, ${V}^{eff}=U_{1}^{eff}$ for $z_{1}<z_{i}<z_{2}$, ${V}^{eff}=U_{2}^{eff}$ for $z_{3}<z_{i}<z_{4}$, ${V}^{eff}=U_{3}^{eff}$ for $z_{5}<z_{i}<z_{6}$ ... 
                    where 
		    \begin{equation}
                    U_{1}^{eff}=- \ln{\left[(1+\exp[{-U_{1}+\mu_{1}}]+
           \exp[{\mu_{2}}]+ \exp[{\mu_{3}}]+...)/ \xi_{0}\right]},
           \end{equation}
            \begin{equation}
             U_{2}^{eff}=- \ln{\left[(1+\exp[{-U_{2}+\mu_{2}}]+
          \exp[{\mu_{1}}]+ \exp[{\mu_{3}}]+...)/ \xi_{0}\right]}
          \end{equation}
and           
              \begin{equation}
             U_{3}^{eff}=- \ln{\left[(1+\exp[{-U_{3}+\mu_{3}}]+
          \exp[{\mu_{1}}]+ \exp[{\mu_{2}}]+...)/ \xi_{0}\right]}.
           \end{equation}
One should note that $z_3 > z_2$, $z_5 > z_4$ and so on, i.e. the binding wells do not overlap.

In the next section, we study the phase behavior of membranes with short and long stickers utilizing the mean field approximation.

\section{Mean-field approximation}
When fluctuation of the membranes  not too strong, one can apply mean-field approximation to the discretized  Laplacian 
$ \Delta_{d}z_i= 4[z_{min} -z_{i}]$ where  $z_{min}$ designates  the average separation field.
         Substituting this equation in 
      Eq. (1), the elastic term takes a simple form:
        $ H[z]= \sum_{i}(4[z_{min} -z_{i}])^2$. Within the mean field approximation, one can write Eq. (5) as  
           \begin{eqnarray}
            Z&=&(\xi_{0})^N
         \left[\int_{0}^{\infty }dz_{i}\exp{[-8(z_{min} -z_{i})^2-{ V}^{eff}(i)]}\right]^N.
           \end{eqnarray}
          The free energy   of the membrane after some algebra is given by  
             \begin{eqnarray}
          G&=&-
    \ln[\left[\int_{0}^{\infty }dz_{i}[\exp[-8(z_{min} -z_{i})^2-{ V}^{eff}(i)] \right] /\xi_{0}].
           \end{eqnarray}
 Varying  the free energy (11) with  respect to $z_{min} $  leads to a self consistence equation of the form:
            \begin{equation}
         z_{min} ={\int_{0}^{\infty }z\exp[-8(z_{min} -z_{i})^2-{ V}^{eff}(i)]dz_{i}\over \int_{0}^{\infty }\exp[-8(z_{min} -z_{i})^2-{ V}^{eff}(i)]dz_{i}}.
           \end{equation}
\begin{figure}[h] 
\begin{center} 
\epsfig{file=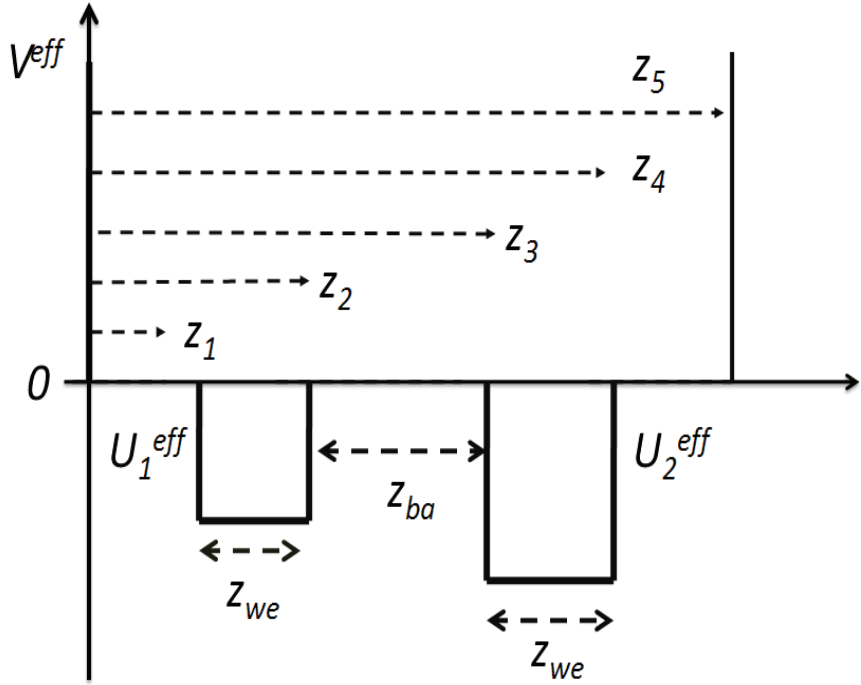,width=12cm}
\caption{Double-well potential ${ V}^{eff}$ with two degenerate minima at both sides of the wells. 
The potential has two square wells of depth ${ U_{1}}^{eff}$ and ${ U_{2}}^{eff}$ within the range 
$z_{we}=z_{2}-z_{1}$ and  $z_{we}=z_{4}-z_{3}$, respectively. 
The effective potential $V^{eff}=\infty$ for $z\le 0$ and $z\ge z_{5}$. We fix $z_{a}=z_{5}-z_{4}=z_{1}$. }
\end{center}    
\end{figure}  

\subsection{Membranes with short and long stickers}

Let us now consider membranes interacting via two types of stickers whose equilibrium phase behavior are governed by an
effective double-well 
potential with the effective depth ${ U}^{eff}_{1}$ and ${ U}^{eff}_{2}$ as  shown in Fig. 2.  ${ U}^{eff}_{1}$ and ${ U}^{eff}_{2}$ depend on the stickers binding energies ${ U}_{1}$ and ${ U}_{2}$, respectively.  Hereafter, all energetic quantities are given in unit of $k_{B}T$ and to 
  make the  model analytically solvable (for symmetry reason) let  $z_{1}=0$,  $z_{4}=z_{5}$,  $z_{we}= z_{2}-z_{1}= z_{4}-z_{3}$ 
and $z_{ba}= z_{3}-z_{2}$.

When membranes are confined in one of the wells of the effective double-well potential, they remain stable with respect to thermal fluctuations provided that the 
 potential wells are deep enough, membranes exhibit two coexisting states with two average separations $ z_{min}$ \cite{h19}. In this case the membranes remain trapped    in one of the wells and exhibit a first order transition. However for weak potential wells, the membrane can easily surmount the potential barrier and its  average separation field  $ z_{min} =0$. 
 
At what critical potential depth ${ U}^{eff}_{C}$ do membranes confined in the double-well potential   $"tunnel"$? This can be addressed  by 
  solving Eq. (12) numerically. For instance  for parameter choice $z_{we}=0.4$, $z_{ba}=0.4$ and $|{ U}^{eff}|=8$, the graphical solution is depicted in Fig. 3. The figure clearly reveals    $z_{min} $ attending two distinct  values  $z_{min}=0.2 $ and $z_{min}=0.98 $  which exhibits   $|{ U}^{eff}|>|{ U}_{C}^{eff}|$. On other hand for  $z_{we}=0.4$, $z_{ba}=0.4$ and $|{ U}^{eff}|= 0.2$, $z_{min} =0$   revealing   $|{ U}^{eff}|<|{ U}_{C}^{eff}|$ (see Fig. 3). 
\begin{figure}
\begin{center}
\epsfig{file=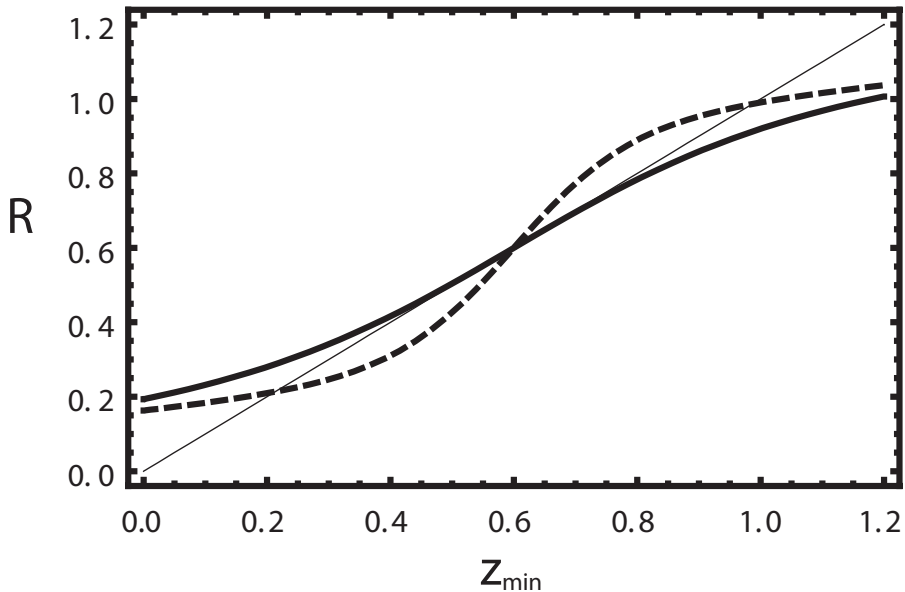,width=12cm}
\caption{  Graphical solution  $ R=\int_{0}^{\infty }z\exp[-8(z_{min} -z_{i})^2-{ V}^{eff}(i)]dz_{i}/ \int_{0}^{\infty }\exp[-8(z_{min} -z_{i})^2-{ V}^{eff}(i)]dz_{i}$   versus $z_{min}$ for  parameter choice  $z_{we}=0.4$ and  $z_{ba}=0.4$.  The dashed and the thick solid  lines  represent  the graphical solution for  $|{ U}^{eff}|=8$  and $|{ U}^{eff}|= 0.2$, respectively.  }
\end{center}
\end{figure}
 \begin{figure}[h] 
\begin{center}
\epsfig{file=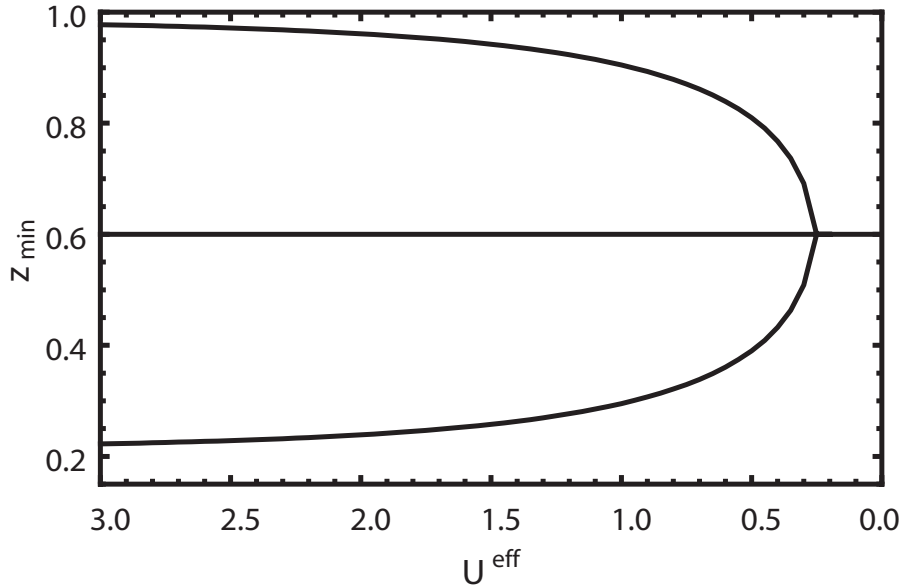,width=12cm}
\caption{ $z_{min} $ versus ${ U}^{eff}$ for fixed $z_{we}=0.4$ and $z_{ba}=0.4$.   The critical point is located at $|{ U}^{eff}_{C}|=0.264$.
For $|{ U}^{eff}|>0.264$, membranes attend two different mean values while   when $|{ U}^{eff}| <0.264$, $z_{min}=0$.}
\end{center}
\end{figure}
Furthermore for parameter values of 
 $z_{we}=0.4 $ and $z_{ba}=0.4$, the dependence of $z_{min}$ as a function of $U^{eff}$ is plotted in  the  Fig. 4. The same figure depicts when $ | U_{}^{eff}|>| U_{C}^{eff}|$,  membranes are confined within one of the potential wells. As the effective potential depth decreases, the tendency for membranes to stay in the potential wells decreases. 
When  $ | U_{}^{eff}|<| U_{C}^{eff}|$, membranes  overcome the barrier height and as the result   $z_{min}=0$.

The critical potential depth ${ U}^{eff}_{C}$,   below which phase separation occurs,  can be obtained by minimizing the free energy, i.e.;
$
-{\partial^2 G/ \partial z_{min} ^2}=0. 
$ 
For fixed $z_{1}=0$ and $z_{4}=z_{5}$, one finds the critical potential depth
\begin{equation}
{ U}_{c}^{eff}=\ln[1-\exp[-8(z_{we}(z_{we}+z_{ba})](1+{2z_{we}\over z_{ba}})].
\end{equation}
Exploiting Eq. (13), one  confirms that when the potential width $z_{we}$
steps up, ${ U}_{C}^{eff}$ declines  which is consistent with the work \cite{h18,h22}. Note that $z_{ba}$ signifies the length difference between the two stickers. As it can be readily seen in Eq. (13), when $z_{ba}$ increases, ${ U}_{C}^{eff}$ declines. This is because when the length difference between the two stickers increases, the energy cost of keeping the short and long stickers in close proximity increases.

The  effective critical  potential depth  ${ U}_{C}^{eff}$ relies on sticker binding energy  ${ U}$ and  stickers chemical potential ${\mu}$. Comparing  Eqs. (6) with  (13)  and assuming ${\mu}_{1}={ \mu}_{2}={ \mu}$, one obtains the  stickers critical binding energy 
\begin{eqnarray}
 U_{C}&=&\mu-\ln[{\beta_{1} \over \beta_{2}}]
\end{eqnarray}
where 
\begin{eqnarray}
 \beta_{1}&=&e^{\mu+8 z_{we}(z_{we}+z_{ba})}z_{ba} +(1+e^{\mu})(z_{ba}+2z_{we})
\end{eqnarray}
 and 
 \begin{eqnarray}
 \beta_{2}&=&-1+e^{8z_{we}(z_{we}+z_{ba})}z_{ba} -2z_{we}.
\end{eqnarray}
When the stickers binding energy  $U_{C}$ is less than the critical binding energy  ${U}<{ U}_{C}$, the short and long stickers segregate into distinct domains while when   ${ U}>{ U}_{C}$, the stickers become mixed.  
The dependence of ${ U}_{C}$ on model parameters can be explored via Eq. (14).  When the length between short and long stickers $z_{ba}$   as well as the width of the stickers potential $z_{we}$ increases, the two types of stickers separate into two distinct domains at smaller critical binding energy ${U}_{C}$. 
 
 Let us now consider a case where $z_{1}\ne 0$, $z_{1}=z_{5}-z_{4}=z_{a}$ 
(the two wells of the  effective double-well potential are far from the hard wall) as shown in Fig. 2. In this case
 the effective critical potential depth ${ U}_{c}^{eff}$  after some algebra is given by  
 \begin{eqnarray}
{ U}_{C}^{eff}&=&\ln[\exp[(8z_{a}m_1)]((\exp{(8z_{we}(z_{we}+z_{ba}))}-1)z_{ba}-\nonumber \\&& 2z_{we})/\alpha]
 \end{eqnarray}
 where 
 \begin{eqnarray}
\alpha&=&z_{ba}+ z_{ba}\exp[(8(z_{a}+z_{we})m_2]-\exp[(8z_{a}m_1)]z_{ba}+\nonumber  \\ &&
2(z_a+z_{we}-\exp[(8z_{a}m_1)]z_{we}).
 \end{eqnarray}
 The values for $m_1$ and $m_2$ are  given by
 $m1=z_{ba}+z_a+2z_{we}$
 and
 $m2=z_{ba}+z_a+z_{we}$. One can note that in the limit $z_{a} \to 0$, Eq. (17) converges to Eq. (13).   
Utilizing  Eq. (17), one can see that as $z_{we}$ and $z_{ba}$ increase, ${ U}_{c}^{eff}$ declines which suggests that as the width of the stickers potential as well as  the length between the short and long stickers increases, the stickers segregate into two distinct domains at a lower stickers critical potential depth.  In the limit $z_{ba}  \to \infty$ or  $z_{we} \to \infty $, $|{ U}_{C}| \to 0$ while in the limit $z_{ba}  \to 0$ or  $z_{we} \to 0 $, $|{ U}_{C}| \to \infty$.  When $z_{a}$ increases,    the entropic repulsion of  membranes with the wall reduces. Thus  the critical potential depth of the stickers ${ U}_{c}^{eff}$   decreases which  exhibits that   confinement facilitates demixing of stickers and hinders  lateral phase separation.

\section{Membrane with short, intermediate and long stickers  }
Let us now  consider membranes with short, intermediate and long stickers near to the substrate. Eliminating the stickers degree of freedom results in membranes interacting with triple-well potential with the effective depth ${ U}^{eff}_{1}$, ${ U}^{eff}_{2}$  and ${ U}^{eff}_{3}$ as  shown in Fig. 5.  ${ U}^{eff}_{1}$, ${ U}^{eff}_{2}$ and ${ U}^{eff}_{3}$    are function of  the stickers binding energies ${ U}_{1}$, ${ U}_{2}$ and ${ U}_{3}$,   respectively.  Once analyzing the effective critical potential depth, one can retrieve the critical point of the stickers binding energy. 
\begin{figure}[h]  
\begin{center}
\epsfig{file=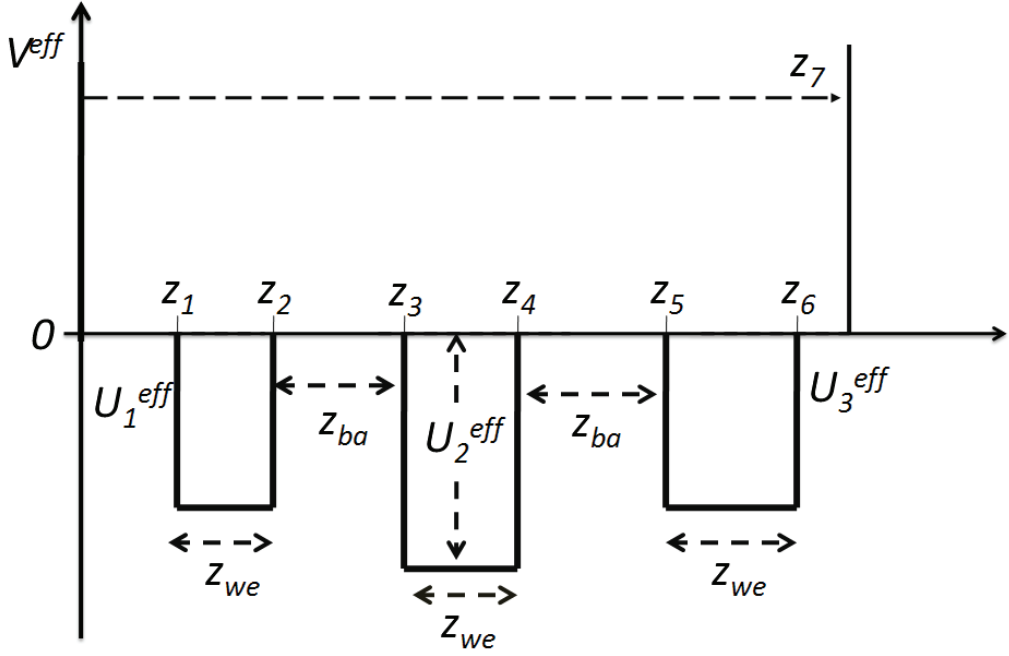,width=14cm}
\caption{Triple-well potential ${ V}^{eff}$ with three degenerate minima. 
The potential has three square wells of depth ${ U_{1}}^{eff}$,  ${ U_{2}}^{eff}$ and ${ U_{3}}^{eff}$ within the range
$z_{we}=z_{2}-z_{1}=z_{4}-z_{3}=z_{6}-z_{5}$, respectively. Here $z_{a}=z_{1}=z_{7}-z_{6}$.
The effective potential $V^{eff}=\infty$ for $z\le 0$ and $z\ge z_{7}$.
}
\end{center}
\end{figure}   

Our earlier analysis unveils  that for membranes with two types of stickers, the  length disparity between the short and long stickers causes lateral phase separation. For pronounced length difference, lateral phase separation occurs  even at high temperature or equivalently at shallow critical potential depth. However the situation is quiet different when there are intermediate stickers embedded  in the membranes. The presence of intermediate stickers reduces the bending energy cost as the length gap between the short and long stickers gets diminished.  Hence in the presence of additional intermediate stickers, the lateral phase separation among the stickers occurs at lower temperature or deeper critical potential depth which suggests that lateral phase separation occurs at the expense of high sticker  binding energy and thus the intermediate stickers endorse  mixing of stickers. 

\begin{figure}[h] 
\begin{center} 
\epsfig{file=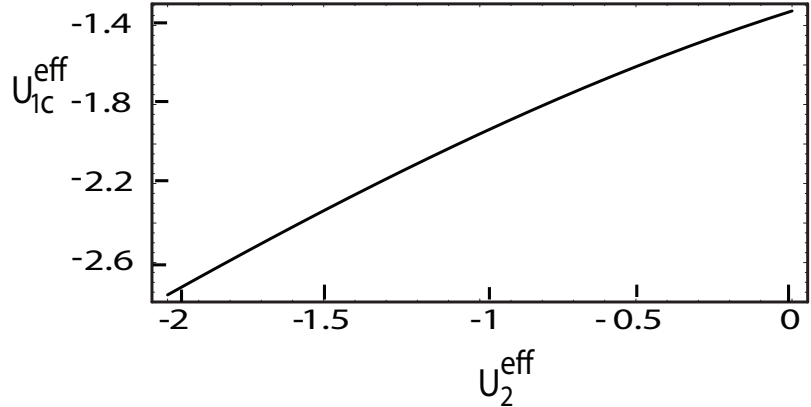,width=12cm}
\caption{The critical  effective potential depth ${ U}_{1C}^{eff}={ U}_{3C}^{eff}$ versus  ${ U}_{2}^{eff}$ for parameter choice  $z_{a}=0.1$, $z_{ba}=z_{we}=0.16$.  }
\end{center}    
\end{figure} 
\begin{figure}[h] 
\begin{center} 
\epsfig{file=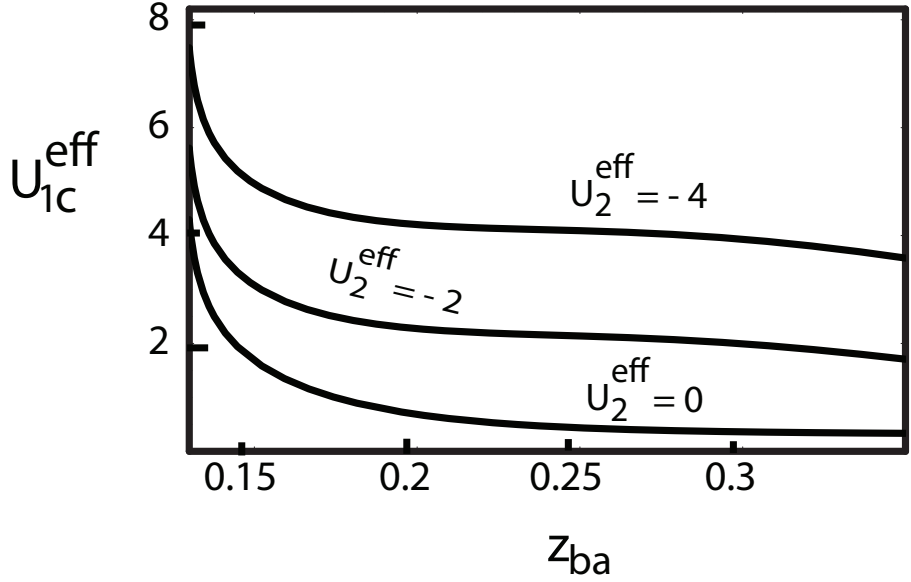,width=12cm}
\caption{The critical  effective potential depth ${ U}_{1C}^{eff}$ versus  $z_{ba}$ for  paranmeter values of  $|{ U}_{2}^{eff}|=4$, $|{ U}_{2}^{eff}|=2$, $|{ U}_{2}^{eff}|=0$, $z_{a}=0.1$ and  $z_{we}=0.16$. }
\end{center}    
\end{figure}

The dependence of the critical potential depth on the model parameters can be explored numerically via   Eq. (12). In order to conceive the effect of the intermediate stickers, let us vary  $U_{2}^{eff}$ and for simplicity let $U_{1}^{eff}=U_{3}^{eff}$.  Figure 6 depicts the plot $U_{1c}^{eff}=U_{3c}^{eff}$ as a function of $U_{2}^{eff}$ for parameter values $z_{a}=0.1$, $z_{ba}=z_{we}=0.16$. When  $U_{2}^{eff}=0$, $|U_{1c}^{eff}|,~|U_{3c}^{eff}|= 1.326$  while as   $|U_{2}^{eff}|$ steps up, $|U_{1c}^{eff}|,~|U_{3c}^{eff}|$ increases revealing the intermediate stickers hindering the lateral phase separation between the short and long stickers. 
Figure 7 exhibits the plot of   $|U_{1c}^{eff}|,~|U_{3c}^{eff}|$ as a function of 
$z_{ba}$ for fixed $U_{1}^{eff}$, $z_{a}=0.1$ and  $z_{we}=0.16$. As $z_{ba}$  increases $|U_{1c}^{eff}|,~|U_{3c}^{eff}|$ declines; when $|U_{2}^{eff}| $ steps up
 $|U_{1c}^{eff}|,~|U_{3c}^{eff}|$ increases once again shows that the intermediate stickers facilitate   mixing of the stickers.

These ideas can easily checked in experiment by considering vesicle adhesion to supported membrane with several types of stickers. Our theoretical prediction reveals that the presence of intermediate stickers suppresses lateral phase separation. This effect can be checked by incorporating different types of adhesive molecule of different size to membrane surface. As the number of species of stickers increases, lateral phase separation gets suppressed.  However lateral phase separation of two or more types of stickers has been observed in experiment. This leads to the conclusion  that either nature keeps its types of stickers  to a limited number or equilibrium modeling, although it is simple, fails to capture the phase behavior of membrane with several types of stickers.

\section{Summary and conclusion}
The equilibrium phase behavior of membranes with stickers of several types  is explored at the mean field level. We first consider membranes  with short and long stickers. We investigate how the critical potential depth $|{ U}_{c}^{eff}|$  behaves as the function of the model parameters. The central result shows that $|{ U}_{c}^{eff}|$ decreases  not only as   the width and  length of the stickers  increase but also when $z_{a}$ increases. In the presence of  intermediate stickers, the numerical results uncover that the intermediate stickers hinder the lateral phase separation.

In  conclusion, via mean field approximation,  we study adhesion induced lateral phase separation of membranes interacting with multiple species stickers of different sizes. Though the mean-field approximation neglects the effect of fluctuations, we believe that the qualitative behavior of the result obtained will not be affected and thus this work is crucial not only for fundamental understanding of the physics of membranes but also for the construction of artificial membranes.

       \section*{\bf Acknowledgement}
I would like to thank Hsuan-Yi Chen for interesting discussions I had during my  visit  at National Central  University,  Department of Physics and Graduate Institute of  Biophysics, Taiwan. I would like also to thank  Thomas Weikl  for his helpful comments,
suggestions and  critical reading of this manuscript. 
It is my pleasure to thank Chun Biu Li and Prof. Tamiki Komatsuzaki for the interesting discussions and for providing a wonderful
research environment.

\end{document}